# Vibro-Thermal Wave Radar: Application of Barker coded amplitude modulation for enhanced low-power vibrothermographic inspection of composites

Saeid Hedayatrasa [1,2,*], Joost Segers[1], Gaétan Poelman[1,2], Wim Van Paepegem[1] and Mathias Kersemans[1]

[1] Mechanics of Materials and Structures (UGent-MMS), Department of Materials, Textiles and Chemical Engineering, Ghent University, Technologiepark-Zwijnaarde 46, 9052 Zwijnaarde, Belgium

[2]SIM Program M3 DETECT-IV, Technologiepark-Zwijnaarde 48, B-9052 Zwijnaarde, Belgium

[*] Corresponding author: saeid.hedayatrasa@ugent.be

**Abstract**
This paper investigates the performance of the thermal wave radar imaging technique in low-power vibrothermography, so-called vibro-thermal wave radar (VTWR), for non-destructive inspection of composites. VTWR is applied by binary phase modulation of the vibrational excitation using a 5 bit Barker coded waveform, followed by matched filtering of the thermal response.

The depth resolvability of VTWR is analyzed by a 1D analytical formulation in which defects are modeled as subsurface heating sources. The obtained results reveal the outperformance of VTWR compared to the classical lock-in vibrothermography (LVT), i.e. a sinusoidal amplitude modulation.

Furthermore, the VTWR technique is experimentally demonstrated on a 5.5 mm thick carbon fiber reinforced polymer coupon with barely visible impact damage. A local defect resonance frequency of a backside delamination is selected as the vibrational carrier frequency. This allows for implementing VTWR in the low-power regime (input power < 1 Watt). It is shown that the Barker coded amplitude modulation and the resultant pulse compression efficiency lead to an increased probing depth, and can fully resolve the deep backside delamination. It is also shown that VTWR approach allows depth-selective imaging of defects through the lag of the compressed pulse.

## 1. Introduction

Active infrared thermography is a cost-effective non-destructive testing technique which enables fast full-field inspection of relatively large objects using a highly sensitive infrared camera [1-3]. The test-piece is generally excited with an external heat source so that a heat flow is stimulated throughout the sample and the defects are detected based on their impact on the thermal response recorded at the inspection surface. A surface heat flux can be induced by irradiating the exterior of the test-piece using e.g. optical lamps or laser (optical thermography). The heat diffuses throughout the material and the defects are detected due to the thermal conductivity mismatch introduced by their interfaces. This needs a double travelling distance of the heat wave to the defect's depth and back (while experiencing a highly damped 3D heat diffusion) which makes relatively deep defects hardly detectable. In fact, the thermal signature of the defect must be sufficiently high so that it dominates the non-uniform heating induced by the excitation source and corresponding in-plane heat diffusion. An alternative approach to overcome this limitation is to inspect the test-piece in the transmission mode such that the defects are detected based on the thermal response transmitted to the back surface [4]. However, this approach requires access to both sides of the test-piece which limits its application for in-situ non-destructive testing (NDT) and health monitoring of structural components. Anyhow, a defect may still be inaccessible due to its poor interaction with the stimulated heat wave, e.g. a closed crack with small effective disbond area or a crack oriented parallel to the heat flow. An alternative to external optical heating, is to induce ohmic heating inside the material by placing the test piece in the proximity of an alternating electromagnetic field, so-called eddy current or induction thermography [5]. The defects that are located in the penetration depth range of the electromagnetic field are detected based on their direct interaction with eddy currents. The deeper defects are further detected through their interaction with the diffusion of heat into the depth. The major limitation of the technique is that its application is limited to the inspection of ferromagnetic materials.

Vibrothermography (also known as sonic thermography or thermosonics) is another active infrared thermography technique in which the test-piece is subjected to an external vibrational excitation e.g. using an actuator bonded to the surface [6]. The dynamic response of the test-piece leads to activation of hardly detectable defects and makes them act as internal heating sources [7, 8]. The vibration-induced heat generated at the defected area directly diffuses to the generally cold inspection surface and reveals the defect when its thermal signature is above the noise level of the infrared camera (i.e. 20 mK for a high-end cooled camera). Although adequate vibrational activation of the defects normally requires very high excitation power of a few kilowatts [9], tuning the excitation frequency band at a local defect resonance (LDR) frequency, enables low-power vibrothermography using a piezoelectric wafer or an air-coupled transducer [10-14]. As LDR frequencies should be known a priori for the LDR based low-power vibrothermography, a more recent study by the current authors [15] has paved the ground for a stand-alone identification of LDR frequencies through an efficient vibrothermographic spectroscopy procedure.

Sinusoidal amplitude modulation (AM) of the heating excitation for a number of cycles, so called lock-in thermography [16], increases the signal-to-noise ratio (SNR). Moreover, the probing depth can be tuned by the AM frequency which controls the diffusion length of the heat wave, i.e. the lower the frequency, the deeper the probing depth. As such the probing depth of lock-in thermography is limited to its fixed AM frequency, which may be extended by considering a broadband frequency or phase modulation of amplitude.

A frequency modulated excitation was initially implemented for broadband thermal wave imaging by Mandelis [17], [18]. Furthermore, it was extended to the concept of thermal wave radar (TWR) [19-22] by adapting the pulse compression technique which was originally developed for increasing the range resolution and the SNR of radio wave radar systems [23]. In TWR, a modulated waveform is used as the excitation signal and its cross-correlation with the corresponding thermal response is calculated. The purpose is to produce an estimate of the impulse response of the sample (considered as a linear and time invariant system) as close as possible to the one obtainable by a pulsed-excitation (Dirac delta-like stimulus), but with a higher SNR. The cross-correlation process compresses the energy of the signal under a main lobe whose peak value and corresponding lag (delay time), respectively, determine the strength of the echo reflected from a subsurface defect and its corresponding depth. Analogue frequency modulated (sweep) and discrete phase modulated (Barker coded) excitations are the two widely researched types of modulated waveforms in TWR [20, 24-28]. Recently, the current authors introduced a novel optimized discrete frequency-phase modulated waveform which outperforms the existing waveforms in terms of depth resolvability [29, 30]. The TWR approach is not exclusive to optical heating, and it has already been applied for enhanced performance and SNR of eddy current infrared thermography [31-34]. In vibrothermography, the concept of sinusoidal amplitude modulation is extensively studied e.g. [10, 14, 35, 36], resulting in lock-in vibrothermography (LVT). Application of TWR in vibrothermography has also been studied by Liu, Gong [37] through linear frequency modulation of the vibrational amplitude for high-power inspection of a metallic test coupon with flat-bottom holes. It was shown that the peak value of the compressed pulse has a higher SNR compared to the phase images obtained through LVT at different frequencies.

In this paper, Vibro-Thermal Wave Radar (VTWR) is introduced which applies discrete binary phase modulation of the vibrational amplitude, using a 5 bit Barker code, followed by application of the matched filter process. From the VTWR, the peak, lag and phase of the compressed pulse is extracted and analyzed. The 5 bit Barker coded VTWR is compared to a 5 cycles LVT of the same AM frequency (i.e. the same excitation energy), and its outperformance in detection of very deep damage features is demonstrated. In section 2, the theory of VTWR is provided and its depth resolvability is analyzed with a 1D analytical model. In section 3, experimental validation is provided for a 5.5 mm thick carbon fiber reinforced polymer (CFRP) coupon with barely visible impact damage (BVID). Various AM frequencies are tested and the deeper probing depth of VTWR is explicitly confirmed. Moreover, it is demonstrated that the VTWR allows for depth-selective imaging of defects through the lag of the compressed pulse. Section 4 formulates the conclusions.

## 2. Theory of vibro-thermal wave radar (VTWR)

Depending on the orientation, asperities, opening and stress state of the defect's interfaces, various heating mechanisms come into play, including rubbing friction, adhesion hysteresis, viscoelastic damping, thermoelastic damping, and also plastic deformation at the crack tips [7, 8, 38, 39]. Among the various mechanisms, rubbing friction and viscoelastic damping predominantly contribute to the vibration-induced heating. Obviously, the frictional heating is exclusively activated in a defected area. However, the viscoelastic damping is present at the areas with a higher strain energy density, i.e. at a defected area due to local defect resonance [14, 15] or at a non-defected area due to global resonance of the test-piece [40]. In this section, a simplified 1D analytical model is used for simulation of the surface thermal response to vibration-induced heating as a subsurface heating source. It is assumed that the amplitude of vibrational response and the resultant vibration-induced heat linearly scale by the amplitude of excitation. This assumption is valid for the case of (Coulomb) frictional heat dissipation and in the absence of both damping losses and contact acoustic nonlinearity at the defect. VTWR is applied by binary phase modulation of the excitation amplitude using a 5 bit Barker code and its depth resolvability is compared with LVT for a CFRP of 5 mm thickness.

### 2.1. Thermal frequency response to subsurface vibration-induced heating sources

As schematically shown in Figure 1(a), a heat source (i.e. defect) is modelled at a depth $h$ of a solid medium. This defect imposes a uniformly distributed heat flux to the corresponding boundary due to a vibration-induced heating. The amplitude of the vibration-induced heating $q_v$ is modulated by an input excitation waveform $S(t)$.

The thermal response of the solid medium along the depth in z-axis, in the absence of internal heating sources and lateral heat dissipation, is governed by the 1D parabolic equation of heat diffusion [41]:

$$\frac{\partial^2 T(z,t)}{\partial z^2} - \frac{1}{\alpha_z}\frac{\partial T(z,t)}{\partial t} = 0 \qquad (1)$$

$$\alpha_z = \frac{k_z}{\rho C_p} \qquad (2)$$

where $T$ is temperature [K], $t$ is time [s], $\alpha_z$ is thermal diffusivity [m²/s], $k_z$ is thermal conductivity [W/m.K], $\rho$ is the density [kg/m³] and $C_p$ is the heat capacity at constant pressure [J/kg.K]. The vibration-induced heat is generated at the defect's depth (i.e. $z = 0$) and the thermal response of the inspection surface $T(z = h, t)$ is calculated by applying the heat dissipation-free boundary conditions given in Figure 1(a).

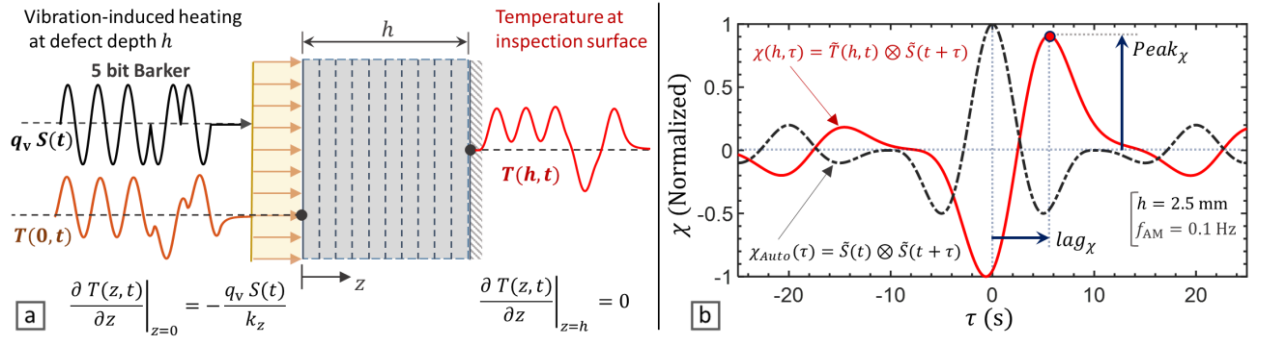

Figure 1: Analysis of the thermal response of a solid medium to subsurface heating sources and implementation of VTWR using a 5 bit Barker coded waveform, (a) schematic model with heat dissipation-free boundary conditions, (b) VTWR for a 2.5 mm deep defect in a CFRP

By assigning the following harmonic solution for steady state thermal response to a mono-frequency excitation:

$$T(z,t) = \theta(z,\omega)\exp(i\omega t) \qquad (3)$$

and substituting in Equation 1, the corresponding thermal frequency response can be calculated as:

$$\theta(z,\omega) = A(\omega)\exp(\beta(\omega)z) + B(\omega)\exp(-\beta(\omega)z) \qquad (4)$$

where $i = \sqrt{-1}$, $\omega = 2\pi f$ is the angular frequency [rad/s], $\beta(\omega) = \sqrt{i\omega/\alpha_z}$, and the constants A and B are derived from the boundary conditions (see Figure 1(a)):

$$A(\omega) = \frac{q(\omega)}{\beta(\omega)k_z}\frac{\exp(-2\beta(\omega)h)}{1-\exp(-2\beta(\omega)h)} \qquad (5)$$

$$B(\omega) = \frac{q(\omega)}{\beta(\omega)k_z}\frac{1}{1-\exp(-2\beta(\omega)h)} \qquad (6)$$

Here, $q(\omega)$ is the frequency-domain heating load which, for a given waveform $S(t)$ and vibration-induced heating amplitude $q_v$, is calculated by:

$$q(\omega) = \mathcal{F}\{q_v S(t)\} \qquad (7)$$

and the corresponding time-domain thermal response at the inspection surface is derived from:

$$T(h,t) = \mathcal{F}^{-1}\{\theta(h,\omega)\} \qquad (8)$$

where $\mathcal{F}$ and $\mathcal{F}^{-1}$, respectively, denote Fourier and inverse Fourier transform operators.

## 2.2. VTWR through matched filtering of the vibro-thermal response

VTWR is implemented by matched filtering of the surface thermal response $T(h, t)$ with the excitation waveform $S(t)$. The matched filter is an optimal linear filter which maximizes the SNR in the presence of stochastic noise and reaches its maximum value at a delay time corresponding to the depth of the defect. Matched filtering is typically done by time-domain cross-correlation as follows [42]:

$$\chi(h, \tau) = \tilde{T}(h, t) \otimes \tilde{S}(t + \tau) = \int_{-\infty}^{+\infty} \tilde{T}(h, t)\tilde{S}(t + \tau)dt \quad (9)$$

where $\otimes$ denotes cross-correlation and (~) denotes the AC component of the signal due to the mono-polar (heating only) nature of vibration-induced heating. In the analytical simulation, a purely harmonic (i.e. bi-polar, or heating-cooling) excitation is applied (i.e. $\tilde{S} = S$ and $\tilde{T} = T$). However, in practice this AC component is estimated by removing a low-order polynomial interpolant of the thermal response as the DC component [25, 29, 43]. For computational efficiency the analysis is performed in the frequency domain as follows [42]:

$$\chi(h, \tau) = \mathcal{F}^{-1}\{\theta(h, \omega)\varsigma^*(\omega)\} \quad (10)$$
$$\varsigma(\omega) = \mathcal{F}\{W(t)\tilde{S}(t)\} \quad (11)$$

where ( * ) denotes the complex conjugate and W is a windowing function used for reducing the side lobes of the cross-correlation. In this study, a Hanning window is applied. The output of cross-correlation $\chi(h, \tau)$ is a sinc-like function which compresses the energy of the whole signal under its main peak as shown in Figure 1(b). The auto-correlation of the excitation signal $\chi_{Auto}(\tau)$ is also included in Figure 1(b) to clarify the asymmetry and the time delay (lag) corresponding to the thermal response from a subsurface defect. The peak value $Peak_\chi$ and corresponding lag $lag_\chi$ of the cross-correlation are then derived as:

$$Peak_\chi = \text{Max}(\chi(h, \tau)) \quad (12)$$
$$lag_\chi = \tau|_{\chi(h,\tau)=Peak_\chi} \quad (13)$$

Subsequently, the phase of cross-correlation $\varphi_\chi$ can be found as:

$$\varphi_\chi = \tan^{-1}\left(\frac{\chi(h, \tau)}{\chi_H(h, \tau)}\right)\bigg|_{\tau=0} \quad (14)$$

where $\chi_H$ is the cross-correlation with the Hilbert transform of $\tilde{S}(t)$ [21]. In the case of mono-frequency harmonic excitation, the phase of cross-correlation $\varphi_\chi$ reduces to the well-known phase of lock-in thermography.

## 2.3. VTWR versus LVT for a CFRP laminate

In this section, VTWR is applied using a broadband 5 bits Barker coded waveform [42] as shown in Figure 1(a). A CFRP with through the thickness thermal conductivity $k_z = 0.53$ W/m.K, density $\rho = 1530$ kg/m$^3$ and specific heat capacity $C_p = 917$ J/kg.K [44] is modelled and $Peak_\chi$, $lag_\chi$ and $\varphi_\chi$ corresponding to subsurface defects in a 5 mm depth range are calculated for two AM frequencies 0.1 and 0.05 Hz, as shown in Figure 2. The 5 bit Barker coded VTWR is compared with a 5 cycles LVT of the same AM frequency, and as such the total excitation energy is the same.

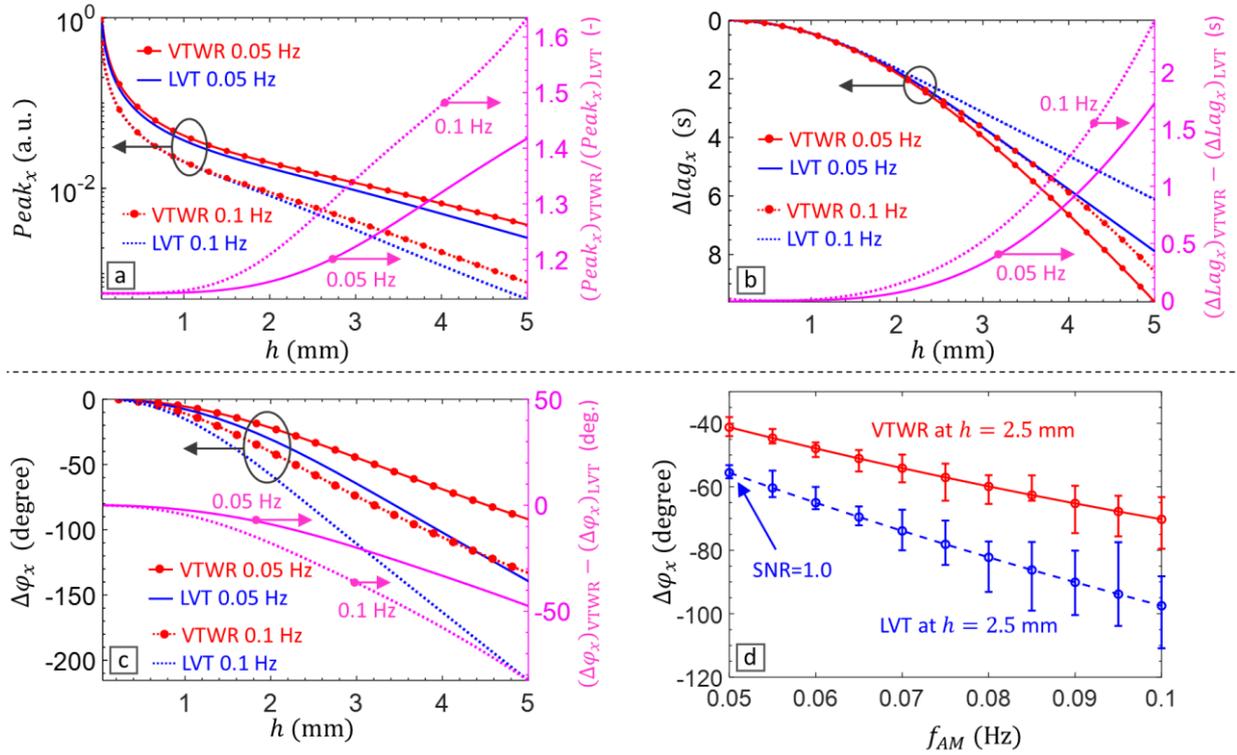

Figure 2: Simulation of VTWR using a 5 bit Barker coded waveform (see Figure 1(a)) and corresponding (a) $Peak_\chi$, (b) $\Delta lag_\chi$, and (c) $\Delta\varphi_\chi$ for inspection of a 5mm thick CFRP, and (d) depth resolvability of $\Delta\varphi_\chi$ at depth $h = 2.5\ mm$ in the AM frequency range 0.05-0.10 Hz and error bars for 20 iterations of adding white Gaussian noise. LVT at the same AM frequency for comparison.

Comparison of the $Peak_\chi$ values (see Figure 2(a), the left axis) indicates the fact that reducing the AM frequency of LVT increases the magnitude of the peak value over the whole depth range, which is further magnified by VTWR. Although the magnitude of $Peak_\chi$ reduces by the depth, its ratio from VTWR to that of LVT (see Figure 2(a), the right axis) rises significantly and at a higher rate for the higher AM frequency. Hence, this amplitude magnification effect, which is gained through the broadband nature of the 5-bit Barker code, increases by the depth of the subsurface heat source and as such enhances the detectability of deeper defects in vibrothermography.

The left axis of Figure 2(b) shows the lag contrast $\Delta lag_\chi$, calculated with respect to the $lag_\chi$ of a very shallow defect (i.e. $h = 0.001$ mm). As expected, $lag_\chi$ increases by the depth and also by lowering the AM frequency. For any individual AM frequency, VTWR has a higher $lag_\chi$ over the depth range. Moreover, the lag resolution from VTWR increases by the depth (see Figure 2(b), the right axis), making VTWR particularly interesting for assessing deep defects. This increased $lag_\chi$ resolution is more pronounced at the higher AM frequency.

Furthermore, the left axis of Figure 2(c) shows the phase contrast $\Delta\varphi_\chi$ with respect to the shallowest defect at $h = 0.001$ mm. Naturally, for both LVT and VTWR the phase contrast increases by depth, but contrary to $lag_\chi$, it reduces by lowering the AM frequency. This can be explained by the fact that although reducing AM frequency from 0.1 to 0.05 Hz doubles the time period of each modulation cycle, the corresponding lag time $lag_\chi$ is only marginally increased as evidenced in Figure 2(b). Therefore, a lower AM frequency results in a lower phase shift at particular depth. Moreover, the broadband modulation introduced by VTWR leads to a lower phase contrast $\Delta\varphi_\chi$ compared to LVT. This is explicitly shown in the right axis of Figure 2(c) which indicates that the degradation of $\Delta\varphi_\chi$ through VTWR is more pronounced at the lower AM frequency and for deeper defects.

In practice, measurement noise degrades the recorded dataset and affects the calculation of $Peak_\chi$, $lag_\chi$ and $\varphi_\chi$. In this respect, it may be expected that the higher $Peak_\chi$ of VTWR leads to a higher SNR, and that its pulse compression efficiency enables a more stable calculation of $lag_\chi$ (this will be shown further in the experimental results of section 3). Therefore, for brevity, the influence of measurements noise is only studied for the calculation of $\varphi_\chi$. For this purpose, the thermal response calculated for a $h = 2.5$ mm deep defect is degraded by adding white Gaussian noise. The noise level is fixed such that the thermal response corresponding to LVT at AM frequency of 0.05 Hz is measured at SNR = 1.0 (i.e., the signal amplitude equals the standard deviation of added noise). The phase contrast $\Delta\varphi_\chi$ is then calculated in the frequency range 0.05 to 0.1 Hz for both LVT and VTWR (see Figure 2(d)). In order to account for the stochastic nature of noise, this procedure has been iterated 20 times and the corresponding variability on the extracted phase contrast $\Delta\varphi_\chi$ is represented by the error bars. An important conclusion of these results is that although VTWR leads to a lower $\Delta\varphi_\chi$ than LVT, it shows lower sensitivity to measurement noise, especially for the higher AM frequency range.

The simulation results based on the simplified analytical solution demonstrate the advantage of a Barker coded VTWR, with respect to LVT of the same AM frequency, in terms of $Peak_\chi$, $lag_\chi$ and $\Delta\varphi_\chi$. Although defect detection through $Peak_\chi$ in optical thermography is quite challenging due to the non-uniform heating induced by the external excitation source, it is of high interest in vibrothermography. This is due to the fact that the vibration-induced heat is distinctively, and almost exclusively, generated at defected areas (particularly at an LDR frequency) and as such is easily identified in the (generally cold) inspection surface.

## 3. Experimental validation of VTWR

In this section the enhanced depth resolvability of vibrothermography through VTWR, as confirmed by the simulation results in section 2, is validated. For this purpose, a $100 \times 150 \times 5.5$ mm$^3$ impacted CFRP coupon, as shown in Figure 3(b,e), with a quasi-isotropic lay-up $[(+45/0/-45/90)_3]_S$ is inspected. The vibrational response of the sample is first measured by scanning laser Doppler vibrometry from which several local defect resonances are identified. An LDR of a deep backside delamination is selected to evaluate the performance of LVT and VTWR at various AM frequencies.

### 3.1. Experimental set-up and LDR selection

The CFRP sample is impacted with a 7.1 kg drop-weight from a height of 0.1 m according to ASTM D7136 [45]. The measured impact energy of 6.3 J introduced Barely Visible Impact Damage (BVID) including a hair-like surface crack at the backside (see the inset of Figure 3(b)). In order to induce broadband vibrational excitation, a low power PZT wafer (type EPZ-20MS64W from Ekulit, with a diameter of 12 mm) is glued to the impact side of the CFRP coupon (using Phenyl salicylate). A Tektronix AFG-3021B arbitrary wave generator together with a Falco System WMA-300 voltage amplifier is then used to supply an excitation voltage of 150 Vpp to the PZT. The mechanical power transmitted to the sample by this set-up is calculated about 200 mW, and as such confirms the low-power levels used in the here described vibrothermography experiments.

In order to study the LDR behavior of the CFRP with BVID, the vibrational response of the sample is measured using a 3D infrared scanning laser Doppler vibrometer (Polytec PSV-500-3D XTRA) in the frequency range (1-250 kHz). The total measurement time of the laser Doppler vibrometry measurement is in the order of 18 minutes. BVID is comprised of a complex combination of damage features through the depth [46], introducing multiple LDRs measured at both the impact side and the backside of the sample [47]. Among the different measured LDRs, an LDR frequency of 91.3 kHz is chosen to be tested by vibrothermography which corresponds to a deep backside delamination in BVID (see Figure 3(a)). At this frequency, only the backside of the sample manifests a prominent in-plane LDR (see the indicated region D on Figure 3(c)). The (practically accessible) impact side (Figure 3(f,g)) is totally transparent to this LDR and merely indicates a global in-plane resonance of the sample. In recent studies by the current authors, the concept of in-plane LDR was introduced [47] and its high efficiency in vibrothermographic NDT of impacted CFRP coupons was demonstrated [14]. Whereas the sample is also experiencing a global in-plane resonance at this LDR frequency, it is expected that the vibrational nodes (e.g. the indicated region N on Figure 3(c,f)), which experience high in-plane strain energy density, heat up due to corresponding high damping losses. It should be noted that all surface maps of Figure 3 are shown with the same colormap scale so that the backside LDR behavior of the BVID at the frequency of 91.3 kHz is distinctively shown. Further, the vibrational response of the backside is mirrored so that the relative location of defects can be conveniently compared.

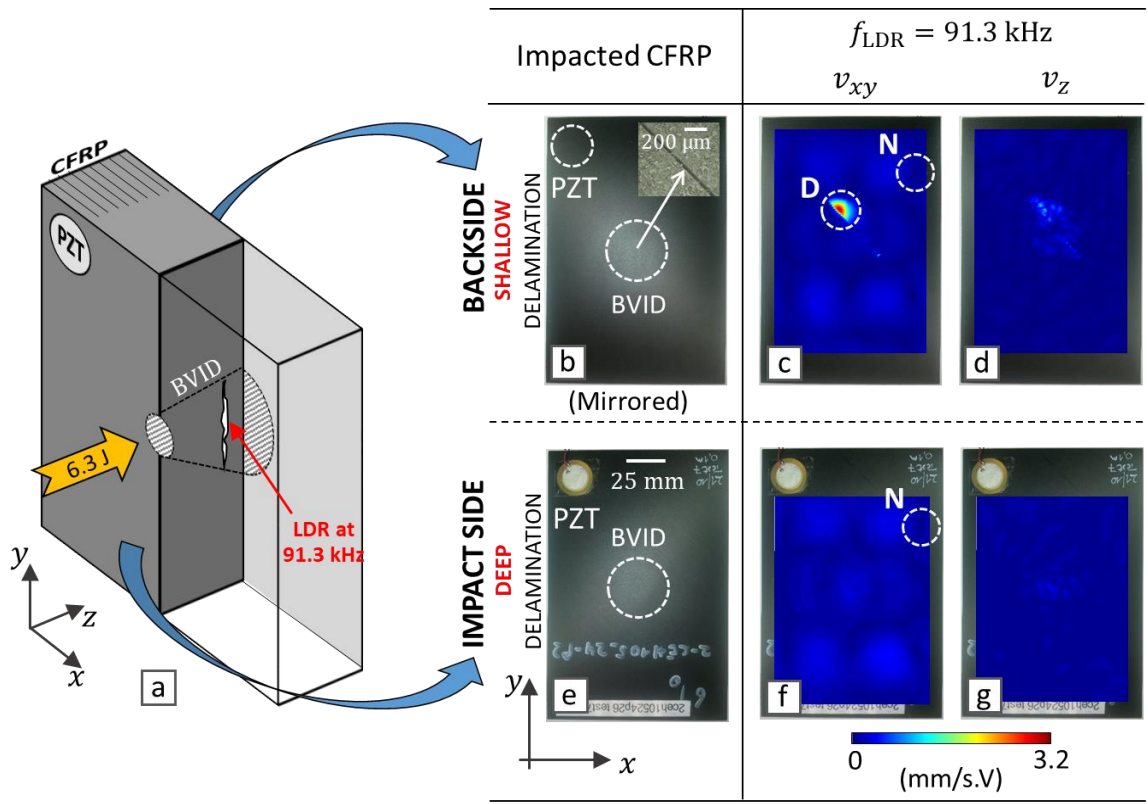

Figure 3: Impacted CFRP coupon of thickness 5.5 mm and corresponding vibrational response (in-plane $v_{xy}$, and out-of-plane $v_z$) measured by 3D infrared scanning laser Doppler vibrometer at a selected LDR frequency $f_{\text{LDR}} = 91.3$ kHz, (a) schematic through-the-thickness presentation of the deep backside delamination, (b-d) backside, (e-g) impact side; (D is the resonated fraction of BVID, and N is an in-plane vibrational node of the CFRP coupon)

For vibrothermography measurements, the CFRP sample is inspected from both impact side and backside, and the depth resolvability in detecting the backside LDR is evaluated. Vibrational excitation is applied at the selected LDR frequency of 91.3 kHz, and is modulated at different AM frequencies 0.1 Hz, 0.075 Hz and 0.05 Hz. VTWR is performed by using a 5 bit Barker coded waveform (see Figure 1(a)) and is compared with LVT of the same duration and energy (i.e. 5 cycles of sinusoidal excitation at the same AM frequency). The surface temperature is measured by a FLIR A6750sc infrared camera (controlled by edevis GmbH hardware-software) at a sampling rate of 25 Hz. The camera has a cryo-cooled InSb detector, a pixel density of $640 \times 512$, a noise equivalent differential temperature (NEDT) of < 20 mK and a bit depth of 14 bit. The output of the infrared camera is given in digital level (DL) scale, which corresponds to the intensity of the emitted infrared radiation. The raw data is exported and further analyzed in MATLAB.

**3.2. VTWR versus LVT at the selected LDR frequency**

Initially the CFRP sample is inspected at the AM frequency of 0.05 Hz. The 5 bit Barker coded waveform is extended with one bit of step heating as shown in Figure 4(a). In this way, the latency of the thermal response to the coded waveform is taken into account and the corresponding AC component is properly decoupled [43]. The AC component is decoupled by removing the DC component as a quadratic polynomial fit of the measured response. The raw temperature and the extracted AC component are shown in Figure 4 for a single pixel picked from the defected area at (b) the backside (shallow delamination) and (c) the impact side (deep delamination).

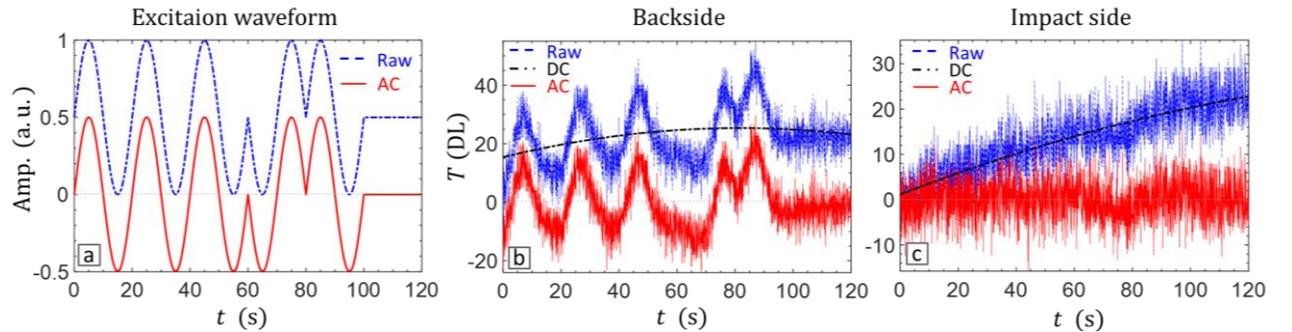

Figure 4: (a) A 5 bit Barker coded waveform extended with one bit of step heating. Raw thermal response and extracted AC and DC components at a pixel in the defected area: (b) backside and (c) impact side.

The surface maps of the calculated $Peak_\chi$, $lag_\chi$ and $\varphi_\chi$ quantities extracted from LVT and VTWR at the AM frequency of 0.05 Hz are shown in Figure 5. The top and the bottom rows correspond to the inspection of the CFRP sample from the backside and the impact side respectively. The $Peak_\chi$ maps of impact side (Figure 5(c,d)) are further post-processed by a median filter with kernel size of 3 pixels.

From the vibrational inspection (see Figure 3), it is clear that the chosen in-plane LDR frequency of 91.3 kHz activates a very deep fraction of the BVID in the backside. For the vibrothermographic inspection of the sample from the backside (see Figure 5(a,b)), the defect is clearly detected with a relatively high peak value, and there is no difference observable between the peak values obtained from LVT and VTWR. By saturating the colormap scale, a few other heating regions (e.g. the region N) are also detected which correspond to the damping losses of in-plane vibrational nodes (see the inset of Figure 5(b)).

For the inspection from the impact side (see Figure 5(c,d)), the same vibrational nodes as well as the very deep defected area D (which was transparent to laser vibrometry from the impact side) are detected. Moreover, the $Peak_\chi$ at defect D is significantly higher for VTWR compared to LVT. Hence, the defect region D is easily discerned from the vibrational node N in the VTWR results. The self-heating of the PZT wafer (attached on the impact side) leads to considerable heat generation, and is therefore saturated to improve the readability of the surface maps.

In terms of $lag_\chi$, the results of both backside and impact side indicate the inefficiency of LVT (Figure 5(e,g)) in detection of heating sources due to its poor pulse compression quality and resultant lag ambiguity in the presence of measurement noise. However, $lag_\chi$ of VTWR (Figure 5(f,h)) provides a meaningful indication of the defected area D and the vibrational nodes. For the backside surface map, the PZT area has the highest $lag_\chi$ indicating that this is the deepest heat source, while the defected area has the lowest $lag_\chi$ indicating that it is the shallowest heat source (and vice versa for the surface map of the impact side). Note that the upper limit of the colormap scale for $lag_\chi$ is set to 20 s to saturate noise and provide better indication of detected features.

In terms of $\varphi_\chi$ (Figure 5(i-l)), the results of both LVT and VTWR provide a clear indication of the vibrational nodes and the defect region D. However, the results of the impact side (Figure 5(k-l)) are significantly different and VTWR provides a less noisy and more distinct indication of heat sources. For consistency, the surface maps of $\varphi_\chi$ show the contrast with respect to the mean noise level of the sound area and the same colormap limits have been applied.

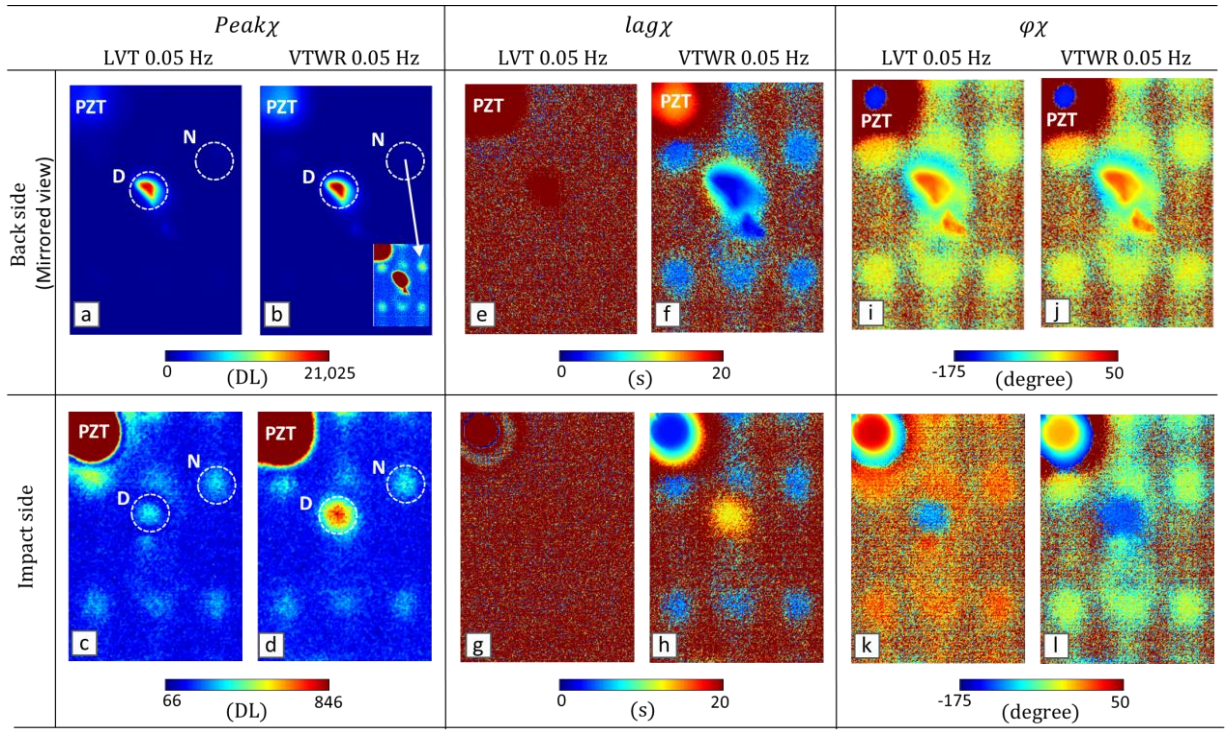

Figure 5: Vibrothermographic inspection of the CFRP sample at LDR frequency 91.3 kHz and AM frequency 0.05 Hz through LVT and VTWR from backside (top row) and impact side (bottom row): (a-d) $Peak_\chi$, (e-h) $lag_\chi$ and (i-l) $\varphi_\chi$. Note that the area heated by the PZT wafer is excluded from the colormap scale in (c) and (d).

The amplitude amplification efficiency of VTWR for inspection of deep defects (herein, inspection of BVID from the impact side) is further demonstrated through the cross-correlation curves of LVT and VTWR averaged over the defected area D (see Figure 6). Application of VTWR on the backside results in a compressed pulse with a prominent peak, which is slightly higher than the amplitude of the sinusoid resultant from LVT. Moreover, application of VTWR on the impact side results again in a compressed pulse, but now with a significantly higher amplitude (more than double) compared to LVT.

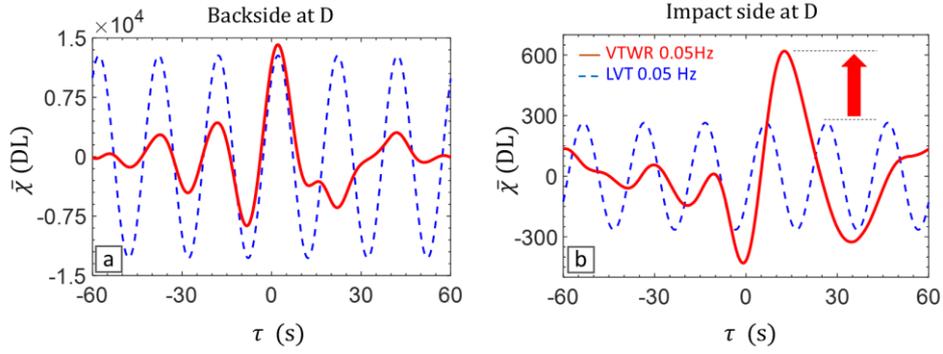

Figure 6: Compressed pulses of LVT and VTWR at AM frequency 0.05 Hz for the defected area D: (a) backside (shallow delamination) and (b) impact side (deep delamination).

### 3.3. Investigation to other AM frequencies

For better understanding the efficiency of VTWR for the detection of deep defects in terms of peak value $Peak_\chi$, the impact side of the sample is further inspected at three different AM frequencies: 0.1, 0.075 and 0.05 Hz. The surface maps of $Peak_\chi$ are shown (only for VTWR) in the top row of Figure 7. Further, a SNR value is calculated for the defected area D and the vibrational node N using the following equation [48]:

$$\text{SNR} = \frac{\left|\overline{Peak_\chi} - \overline{Peak_{\chi S}}\right|}{\sigma_S} \qquad (15)$$

where $\overline{Peak_\chi}$ and $\overline{Peak_{\chi S}}$ are the average values of $Peak_\chi$ over an area of interest (defect D and node N) and the reference sound area S, respectively. $\sigma_S$ is the standard deviation of $Peak_\chi$ over the reference sound area S. The SNR values for both VTWR and LVT are given in the bottom row of Figure 7.

At the highest AM frequency of 0.1 Hz, there is a minor indication of the vibrational nodes and the defected region D (Figure 7(a)). Upon lowering the AM frequency to 0.075 Hz (Figure 7(b)) and further to 0.05 Hz (Figure 7(c)), the defected region D is detected with a considerably higher magnitude compared to the vibrational nodes. The SNRs of LVT and VTWR further confirm the higher magnification efficiency of VTWR for the deep defects (see Figure 7(d-f)). At all AM frequencies, application of VTWR leads to an increased SNR value at the vibrational node N. But more importantly, it increases the SNR of the defected region D with a higher rate. This leads to a distinct detectability of the (deep) backside defect D at the two lower AM frequencies 0.075 Hz and 0.05 Hz. At these frequencies, VTWR allows for distinguishing the heating induced by the defected area D from the 'misleading' heating induced by a non-defected vibrational node like N. This is in clear contrast with the results obtained through LVT: even at the lowest AM frequency of 0.05 Hz, the SNR of the defect region D is at a similar level as the SNR of the vibrational node N.

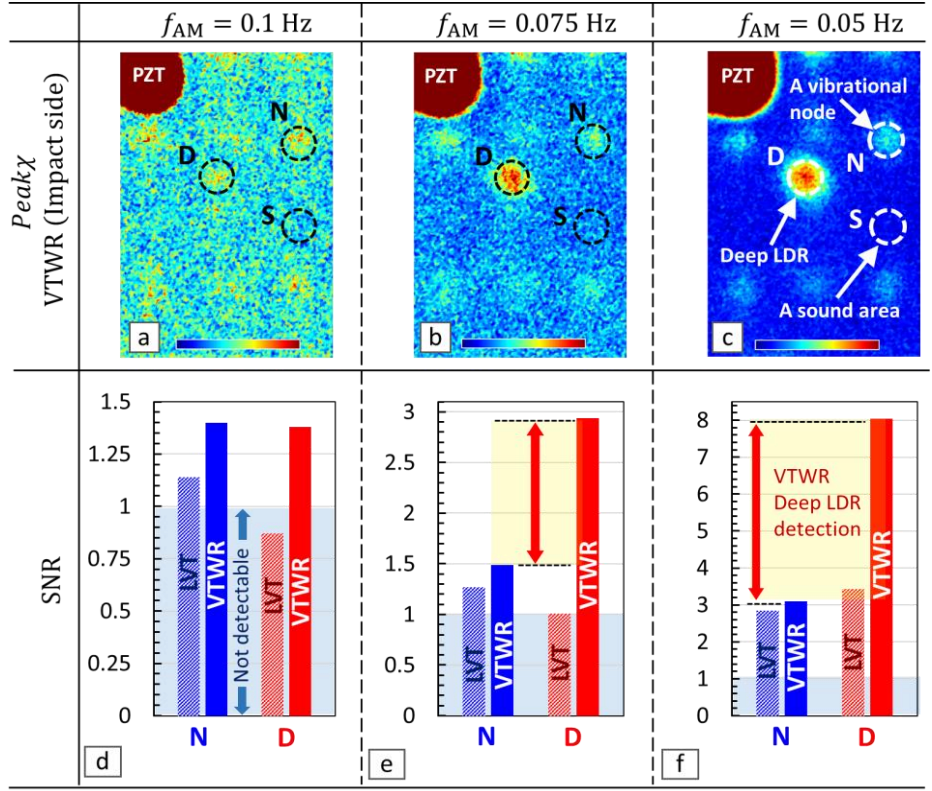

Figure 7: Inspection of the impact side of the CFRP sample at various AM frequencies: (left) 0.1 kHz, (middle) 0.075 kHz and (right) 0.05 kHz. The top row represents the surface maps of $Peak_\chi$ resulting from VTWR, the bottom row displays the SNR values of regions D and N calculated for both LVT and VTWR.

### 3.4. VTWR for selective imaging and depth slicing of heat sources

Considering that the VTWR with Barker coded amplitude modulation provides high pulse compression, and stable indication of $lag_\chi$, it enables selective imaging of the (vibration-induced) heat sources through the thickness of the sample. Such vibro-thermal tomography is done by evaluating the full-field cross-correlation images across the cross-correlation time $\tau$. Individual heat sources are then distinctively visualized at their corresponding lag times $lag_\chi$ of the compressed pulse. This vibro-thermal tomography procedure is demonstrated in Figure 8, in which the surface maps are displayed for different lag times.

The top row of Figure 8 corresponds to the backside of the sample in which the (shallow) defected area D reaches its peak value at the shortest $lag_\chi$ of 2.04 s (Figure 8(b)), the vibrational node at an intermediate $lag_\chi$ of 4.84 s (Figure 8(c)), and the PZT (attached to the impact side) at the highest $lag_\chi$ of 14.36 s (Figure 8(d)). In fact, this high $lag_\chi$ of the PZT area corresponds to the full through-the-thickness transmission of the heat wave (in addition to the electro-thermal latency in self-heating of the PZT).

The bottom row of Figure 8 corresponds to the impact side of the sample in which the (shallow) PZT area reaches its peak value at the shortest $lag_\chi$ of 3.44 s (Figure 8(f)), the vibrational node at the intermediate $lag_\chi$ of 5.04 s (Figure 8(g)), and the (very deep) defected area D at the highest $lag_\chi$ of 12.60 s (Figure 8(h)). This $lag_\chi$ of the defected area D is comparable to that of the PZT when inspecting from the backside (14.36 s), which confirms that it corresponds to a very deep defect.

The surface map corresponding to $lag_\chi$ of 5.04 s at N (Figure 8(g)) merely indicates the viscoelastic heating at the vibrational nodes. The surface map corresponding to $lag_\chi$ of 12.60 s at the defected area D (Figure 8(h)) provides an exclusive indication of this deep defect.

In the supplementary data, the vibro-thermal tomographic procedure is explicitly demonstrated by an animation which slices through the full depth of the impacted CFRP sample. Further calibration of $lag_\chi$ versus depth (e.g. with the aid of 3D finite element simulation) will enable depth quantification of internal heat sources.

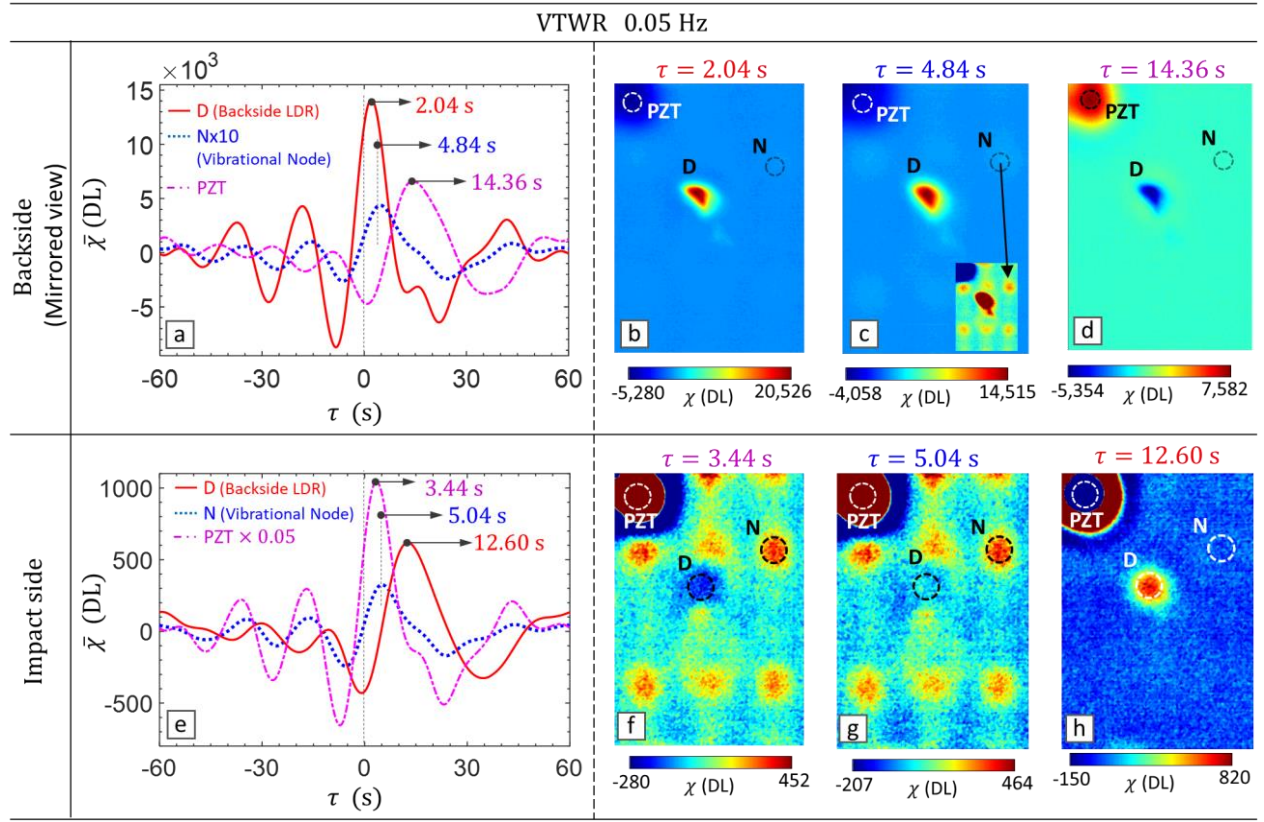

Figure 8: VTWR for selective imaging of heat sources from backside (top row) and impact side (bottom row): (a,e) compressed pulses at selected regions D, N and PZT and (b-d,f-h) full-field maps at corresponding lag times $lag_\chi$ (the area heated by the PZT wafer is excluded from the colormap scale in (f-h))

## 4. Conclusions

In this paper the superior performance of vibro-thermal wave radar (VTWR), compared to classical lock-in vibrothermography (LVT), in detecting deep damage in composites was demonstrated. VTWR was applied by binary phase modulation of the vibrational excitation using a 5 bit Barker coded waveform, and evaluating its cross-correlation with the resultant thermal response. Different amplitude modulation (AM) frequencies were evaluated and the superior through-the-depth resolvability of VTWR to lock-in vibrothermography, in terms of peak and corresponding lag time and phase of cross-correlation, was confirmed.

The thermal response to subsurface heating sources (i.e. vibration-induced heating at defects) was simulated using a simplified analytical model. It was shown that the broadband amplitude modulation applied in VTWR results in a magnification effect on the thermal signature, which is most pronounced for deep defects. Moreover, the outperformance of VTWR in terms of lag and phase was analytically presented.

The analytical results were further validated by vibrothermographic inspection of an impacted CFRP coupon of thickness 5.5 mm. The vibrational response of the sample was measured by scanning laser Doppler vibrometry and a prominent in-plane local defect resonance (LDR) frequency associated with a backside delamination was selected. The sample was then inspected from both sides by vibrothermography at the selected LDR frequency. This backside LDR (which is transparent to laser vibrometry from the impact side) was clearly detected by VTWR, clearly showing it potential for detecting very deep defects in challenging materials.

Finally, by proper analysis of the lag time of the compressed pulse, it was further demonstrated that the VTWR approach allows depth-slicing through the sample for selective defect imaging. This does not only improve the defect detectability significantly, it also allows for vibrothermal tomographic inspection of the composite from which the defect depth can be estimated.

## Acknowledgement

The authors acknowledge the SBO project DETECT-IV (Grant no. 160455), which fits in the SIM research program MacroModelMat (M3) coordinated by Siemens (Siemens Digital Industries Software, Belgium) and funded by SIM (Strategic Initiative Materials in Flanders) and VLAIO (Flemish government agency Flanders Innovation & Entrepreneurship). The authors also acknowledge Fonds voor Wetenschappelijk Onderzoek Vlaanderen (FWO-Vlaanderen) through grants 1S11520N, 12T5418N and 1148018N. The authors express their gratitude towards Honda R&D Co. for supplying material for this research.

## References

1. Yang, R. and Y. He, *Optically and non-optically excited thermography for composites: A review.* Infrared Physics & Technology, 2016. **75**: p. 26-50.
2. Khodayar, F., S. Sojasi, and X. Maldague, *Infrared thermography and NDT: 2050 horizon.* Quantitative InfraRed Thermography Journal, 2016. **13**(2): p. 210-231.
3. Ciampa, F., et al., *Recent advances in active infrared thermography for non-destructive testing of aerospace components.* Sensors, 2018. **18**(2): p. 609.
4. Maierhofer, C., et al., *Characterizing damage in CFRP structures using flash thermography in reflection and transmission configurations.* Composites Part B: Engineering, 2014. **57**: p. 35-46.
5. Wilson, J., et al., *Modelling and evaluation of eddy current stimulated thermography.* Nondestructive Testing and Evaluation, 2010. **25**(3): p. 205-218.
6. Reifsnider, K., E.G. Henneke, and W. Stinchcomb, *The mechanics of vibrothermography*, in *Mechanics of nondestructive testing*. 1980, Springer. p. 249-276.
7. Renshaw, J., et al., *The sources of heat generation in vibrothermography.* NDT & E International, 2011. **44**(8): p. 736-739.
8. Rizi, A.S., et al., *FEM modeling of ultrasonic vibrothermography of a damaged plate and qualitative study of heating mechanisms.* Infrared Physics & Technology, 2013. **61**: p. 101-110.
9. Holland, S.D., et al., *Quantifying the vibrothermographic effect.* NDT & E International, 2011. **44**(8): p. 775-782.
10. Solodov, I. and G. Busse, *Resonance ultrasonic thermography: Highly efficient contact and air-coupled remote modes.* Applied Physics Letters, 2013. **102**(6): p. 061905.
11. Solodov, I., et al., *Highly-efficient and noncontact vibro-thermography via local defect resonance.* Quantitative InfraRed Thermography Journal, 2015. **12**(1): p. 98-111.
12. Fierro, G.P.M., et al., *Nonlinear ultrasonic stimulated thermography for damage assessment in isotropic fatigued structures.* Journal of Sound and Vibration, 2017. **404**: p. 102-115.
13. Dyrwal, A., M. Meo, and F. Ciampa, *Nonlinear air-coupled thermosonics for fatigue micro-damage detection and localisation.* NDT & E International, 2018. **97**: p. 59-67.

14. Segers, J., et al., *In-plane local defect resonances for efficient vibrothermography of impacted carbon fiber-reinforced polymers (CFRP).* NDT & E International, 2019. **102**: p. 218-225.
15. Hedayatrasa, S., et al., *Vibrothermographic spectroscopy with thermal latency compensation for effective identification of local defect resonance frequencies of a CFRP with BVID.* NDT & E International, 2020. **109**: p. 102179.
16. Wu, D. and G. Busse, *Lock-in thermography for nondestructive evaluation of materials.* Revue Générale de Thermique, 1998. **37**(8): p. 693-703.
17. Mandelis, A., *Time-delay-domain and pseudorandom-noise photoacoustic and photothermal wave processes: a review of the state of the art.* IEEE transactions on ultrasonics, ferroelectrics, and frequency control, 1986. **33**(5): p. 590-614.
18. Mandelis, A., *Frequency modulated (FM) time delay-domain thermal wave techniques, instrumentation and detection: a review of the emerging state of the art in QNDE applications*, in *Review of progress in quantitative nondestructive evaluation*. 1987, Springer. p. 799-806.
19. Mulaveesala, R., J.S. Vaddi, and P. Singh, *Pulse compression approach to infrared nondestructive characterization.* Review of Scientific Instruments, 2008. **79**(9): p. 094901.
20. Tabatabaei, N. and A. Mandelis, *Thermal-wave radar: A novel subsurface imaging modality with extended depth-resolution dynamic range.* Review of Scientific Instruments, 2009. **80**(3): p. 034902.
21. Tabatabaei, N. and A. Mandelis, *Thermal coherence tomography using match filter binary phase coded diffusion waves.* Physical review letters, 2011. **107**(16): p. 165901.
22. Tabatabaei, N., A. Mandelis, and B.T. Amaechi, *Thermophotonic radar imaging: An emissivity-normalized modality with advantages over phase lock-in thermography.* Applied Physics Letters, 2011. **98**(16): p. 163706.
23. Nathanson, F.E., J.P. Reilly, and M.N. Cohen, *Radar design principles-Signal processing and the Environment.* NASA STI/Recon Technical Report A, 1991. **91**.
24. Ghali, V., S. Panda, and R. Mulaveesala, *Barker coded thermal wave imaging for defect detection in carbon fibre-reinforced plastics.* Insight-Non-Destructive Testing and Condition Monitoring, 2011. **53**(11): p. 621-624.
25. Gong, J., et al., *Investigation of carbon fiber reinforced polymer (CFRP) sheet with subsurface defects inspection using thermal-wave radar imaging (TWRI) based on the multi-transform technique.* NDT & E International, 2014. **62**: p. 130-136.
26. Laureti, S., et al., *Comparative study between linear and non-linear frequency-modulated pulse-compression thermography.* Applied Optics, 2018. **57**(18): p. D32-D39.
27. Shi, Q., et al., *Study on the Detection of CFRP Material with Subsurface Defects Using Barker-Coded Thermal Wave Imaging (BC-TWI) as a Nondestructive Inspection (NDI) Tool.* International Journal of Thermophysics, 2018. **39**(8): p. 92.
28. Shi, Q., et al., *Barker-coded Modulation Laser Thermography for CFRP Laminates Delamination Detection.* Infrared Physics & Technology, 2019. **98**: p. 55-61.
29. Hedayatrasa, S., et al., *Performance of frequency and/or phase modulated excitation waveforms for optical infrared thermography of CFRPs through thermal wave radar: A simulation study.* Composite Structures, 2019: p. 111177.
30. Hedayatrasa, S., et al., *Novel discrete frequency-phase modulated excitation waveform for enhanced depth resolvability of thermal wave radar.* Mechanical Systems and Signal Processing, 2019. **132**: p. 512-522.
31. Yang, R. and Y. He, *Pulsed inductive thermal wave radar (PI-TWR) using cross correlation matched filtering in eddy current thermography.* Infrared Physics & Technology, 2015. **71**: p. 469-474.

32. Yang, R., et al., *Induction Infrared Thermography and Thermal-Wave-Radar Analysis for Imaging Inspection and Diagnosis of Blade Composites.* IEEE Transactions on Industrial Informatics, 2018. **14**(12): p. 5637-5647.
33. Yi, Q., et al., *New features for delamination depth evaluation in carbon fiber reinforced plastic materials using eddy current pulse-compression thermography.* NDT & E International, 2019. **102**: p. 264-273.
34. Yi, Q., et al., *Quantitative Evaluation of Crack Depths on Thin Aluminum Plate using Eddy Current Pulse-Compression Thermography.* IEEE Transactions on Industrial Informatics, 2019.
35. Rantala, J., D. Wu, and G. Busse, *Amplitude-modulated lock-in vibrothermography for NDE of polymers and composites.* Research in Nondestructive Evaluation, 1996. **7**(4): p. 215-228.
36. Dillenz, A., G. Busse, and D. Wu. *Ultrasound lock-in thermography: feasibilies and limitations.* in *Diagnostic Imaging Technologies and Industrial Applications.* 1999. International Society for Optics and Photonics.
37. Liu, J., et al., *Study of inspection on metal sheet with subsurface defects using linear frequency modulated ultrasound excitation thermal-wave imaging (LFM-UTWI).* Infrared Physics & Technology, 2014. **62**: p. 136-142.
38. Vaddi, J.S. and S.D. Holland. *Identification of heat source distribution in vibrothermography.* in *AIP Conference Proceedings.* 2014. AIP.
39. Truyaert, K., et al., *Theoretical calculation of the instantaneous friction-induced energy losses in arbitrarily excited axisymmetric mechanical contact systems.* International Journal of Solids and Structures, 2019. **158**: p. 268-276.
40. Rodriguez, S., A. Meziane, and C. Pradère, *Thermal Chladni plate experiments to reveal and estimate spatially dependent vibrothermal source.* Quantitative InfraRed Thermography Journal, 2019. **16**(2): p. 163-171.
41. Hahn, D.W. and M.N. Özisik, *Heat conduction.* 2012: John Wiley & Sons.
42. Mahafza, B.R., *Radar Systems Analysis and Design Using MATLAB Third Edition.* 2016: Chapman and Hall/CRC.
43. Silipigni, G., et al., *Optimization of the pulse-compression technique applied to the infrared thermography nondestructive evaluation.* NDT & E International, 2017. **87**: p. 100-110.
44. Maierhofer, C., et al., *Evaluation of Different Techniques of Active Thermography for Quantification of Artificial Defects in Fiber-Reinforced Composites Using Thermal and Phase Contrast Data Analysis.* International Journal of Thermophysics, 2018. **39**(5): p. 61.
45. D7136/D7136M-15, A., *Standard test method for measuring the damage resistance of a fiber-reinforced polymer matrix composite to a drop-weight impact event.* 2015, ASTM International West Conshohocken, PA.
46. Spronk, S.W.F., et al., *Comparing damage from low-velocity impact and quasi-static indentation in automotive carbon/epoxy and glass/polyamide-6 laminates.* Polymer Testing, 2018. **65**: p. 231-241.
47. Segers, J., et al., *Towards in-plane local defect resonance for non-destructive testing of polymers and composites.* NDT & E International, 2018. **98**: p. 130-133.
48. Madruga, F.J., et al., *Infrared thermography processing based on higher-order statistics.* NDT & E International, 2010. **43**(8): p. 661-666.